# Distributed Raman optical amplification in phase coherent transfer of optical frequencies

C. Clivati[1,2], G. Bolognini[3], D. Calonico[1*], S. Faralli[4], F. Levi[1], A. Mura[1] and N. Poli[5]

[1]Istituto Nazionale di Ricerca Metrologica INRIM — strada delle Cacce 91, 10135, Torino, Italy
[2]Politecnico di Torino — Corso Duca degli Abruzzi 24,10129, Torino, Italy
[3]CNR, IMM Inst, Via Gobetti, 101, 40129 Bologna, Italy
[4]Scuola Superiore Sant'Anna, TeCIPInstitute, Via G. Moruzzi 1, 56124 Pisa, Italy
[5]Dip. Fisica e Astronomia, INFN and LENS, Università di Firenze, Via Sansone 1, 50019 Sesto Fiorentino(FI), Italy
*Corresponding author: d.calonico@inrim.it

We describe the application of Raman Optical-fiber Amplification (ROA) for the phase coherent transfer of optical frequencies in an optical fiber link. ROA uses the transmission fiber itself as a gain medium for bi-directional coherent amplification. In a test setup we evaluated the ROA in terms of on-off gain, signal-to-noise ratio, and noise added to the carrier. We transferred a laser frequency in a 200 km optical fiber link with an additional 16 dB fixed attenuator (equivalent to 275 km of fiber on a single span), and evaluated both co-propagating and counter-propagating amplification pump schemes, demonstrating nonlinear effects limiting the co-propagating pump configuration. The frequency at the remote end has a fractional frequency instability of $\sigma_y(\tau) = 3\times10^{-19}$ over 1000 s with the optical fiber link noise compensation.

Nowadays optical frequency standards are outperforming the primary cesium fountain atomic clocks, in terms of stability and accuracy, and are the most promising candidates for a redefinition of the second [1]. This improvement in accuracy and stability would be fruitless without a suitable method to compare and disseminate such ultra-stable optical frequencies over long distances. It has been demonstrated that phase coherent optical fiber links can be used to this purpose [2-4], provided that phase noise introduced by the fiber and its losses are compensated. In this technique light has to travel exactly the same path in both directions, with almost perfect bidirectionality, otherwise the noise cancellation would be ineffective [5]. To overcome the losses and preserve the bi-directionality, special bidirectional Erbium Doped Fiber Amplifiers (BEDFA) [3,4,6] and Fiber Brillouin Amplifiers (FBA) have been developed and are currently used [4,7].

In this Letter, we report on what we believe is the first application of Raman Optical-fiber Amplification (ROA) for time-frequency metrology, in a laboratory optical fiber link with bi-directional phase-noise compensation.

ROA is based on stimulated Raman scattering in silica fibers, a nonlinear effect generating a power transfer between two optical beams, the pump and the signal, through scattering with optical phonons [8,9]. The gain is maximum when the signal light is downshifted ~13.2 THz from the pump light in silica fibers, and has a large bandwidth of several THz around the gain peak. ROA also shows a significant polarization dependence [10], so pump depolarization techniques are commonly employed in standard fiber links. As in FBA, ROA has the most relevant advantage of withstanding very high gain, with no onset of lasing or oscillations [9], thanks to distributed gain along the fiber, whereas in BEDFA Rayleigh scattering causes saturation of the gain medium and the onset of such effects [6,8], compelling to keep the gain to less than 20 dB. Compared with FBA, ROA is intrinsically bidirectional and amplifies both co- and counter-propagating signals, whereas FBA only amplifies signals counter-propagating to the pump. Moreover, ROA has a large gain bandwidth (few THz in silica fibers) so that pump wavelength stabilization is not required to ensure stable gain as in FBA, due to the narrow gain bandwidth (tens of MHz) [7].

ROA has been employed in long-haul fiber-optic transmission in recent years [9,11], but to our knowledge it has not yet been employed in phase-coherent time/frequency metrology applications as an alternative to BEDFA or FBA, and no experimental study of added noise (phase and amplitude) in bi-directional phase-compensated optical links has been performed.

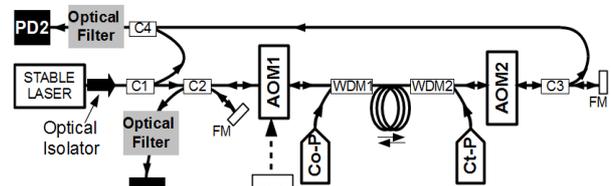

Fig. 1. Set up for ROA in a phase-coherent fiber link. PD Photodiode, C Couplers, FM Faraday Mirror, WDM Wavelength Division Multiplexer, Co-P Co-propagating Pump; Ct-P Counter-propagating Pump, AOM Acousto-Optic Modulator, PLL Phase Locked Loop.

The study involves an evaluation of the gain, the signal-to-noise ratio (SNR) and the added phase noise. We demonstrate the use of ROA in a single-pump scheme on 200 km, then in a double-pump scheme on ~60 dB optical losses, equivalent to 275 km total fiber length, without intermediate amplification. Finally, we point out nonlinearities on the signal for non-optimized ROA gain.

In designing our ROA scheme, care has been taken to avoid the introduction of unwanted noise in the amplified signal, particularly in relation to Kerr-based nonlinear effects, transfer of relative intensity noise (and, to a smaller extent, phase noise) from pump to signal, and double Rayleigh scattering. Such effects, critical in optical transmission systems with signal bandwidths exceeding 40 GHz, are expected to bring a smaller impact on sources used in metrology exhibiting quasi-monochromatic spectra [8].

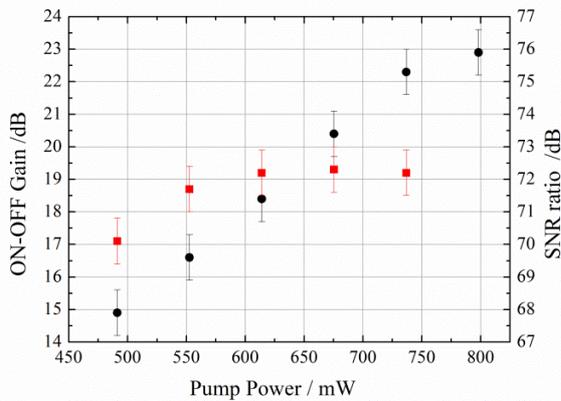

Fig. 2. ROA on-off gain (circles) and SNR (squares) changing counter-propagating pump power.

To study the ROA in a coherent optical link, we set up a laboratory test-bed whose scheme is shown in Figure 1. An ultrastable laser radiation at 1542 nm is generated locking a fiber laser by Pound-Drever-Hall technique to a Fabry-Perot high finesse cavity (120,000) made of Corning Ultra Low Expansion (ULE) glass, ensuring a laser linewidth smaller than 30 Hz (details in [12]). The laser light is coupled into a standard single mode fiber (SMF-28), with a total length up to 200 km, simulating a long-haul link from local to remote laboratory. The light power at fiber input is 0.5 mW; total losses due to fiber and connectors amount to ~45 dB. To achieve laser noise cancellation, two acousto-optic modulators (AOM) are employed. AOM1 is placed before the fiber link and used as actuator to compensate the fiber phase noise. AOM2 is placed after the fiber link, shifts the back-reflected radiation to distinguish between the real signal (shifted by AOM2) and the backscattering from the fiber (not shifted). Photodiode PD1 is employed to detect the phase error from an interferometric scheme [12]; then, a phase-locked-loop (PLL) feeds AOM1 with the correction signal. Photodiode PD2 is used to assess the link performance, comparing the phase noise of the signal at the remote end to that injected at the input by their beatnote.

ROA co-propagating and counter-propagating respect to the transmitted signal was implemented coupling the pump into the fiber through Wavelength Division Multiplexers WDM1 or WDM2 respectively. Two different laser pumps were used. The first was a depolarized fiber laser at 1450 nm delivering output power of up to ~800 mW. The other pump was composed of two polarization-multiplexed Fabry-Perot diode lasers at 1450 nm, providing a depolarized pump with an output power of ~260 mW [13].

We initially evaluated the ROA impact with a counter-propagating pump scheme, using the fiber laser as pump; on-off gain and SNR were measured using the link as a Mach-Zender interferometer, in which the amplified signal was in the measurement arm. We detected on PD2 the heterodyne beatnote between the amplified signal and the light from the short reference arm. Figure 2 reports measured gain and SNR in a 3 kHz bandwidth after 100 km versus pump power coupled into the fiber (1.3 dB insertion loss due to WDM). The gain increased with pump power and attained the maximum value of 23 dB, limited by the effective power available with our pump (800 mW). At high optical pump power, SNR was flat as expected, due to the amplified spontaneous emission [8]. For low optical power (pump radiation lower than 600 mW) the measured SNR showed a slight degradation due to the electrical noise on the photodiode. In these conditions, we measured the phase noise density $S_\phi(f)$ and fractional frequency instability $\sigma_y(\tau)$ of the 100 km optical link with and without optical amplification, to assess possible effects of ROA on the link phase noise.

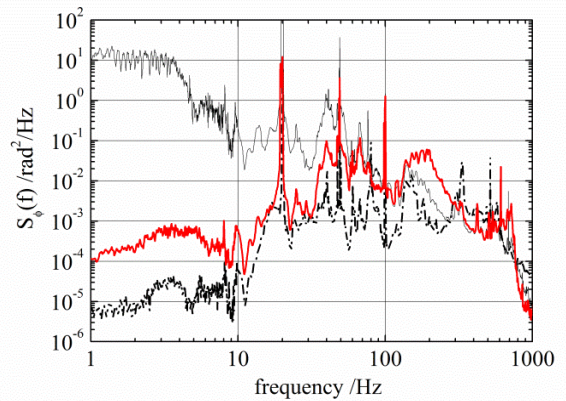

Fig. 3. Phase noise density $S_\phi(f)$ for the compensated link, at 100 km (dashed curve) and 200 km (solid), and free-running link at 200 km (uppermost thin curve).

We did not observe any degradation of phase noise on the compensated link, thus we increased the fiber length to 200 km. Figure 3 reports the closed-loop phase noise at 100 km (lowest dashed curve), at 200 km (second solid curve), and also the phase noise for the uncompensated link at 200 km (uppermost thin curve).The closed-loop noise increase from 100 km to 200 km at frequencies much lower than the locking bandwidth is in good agreement with the theoretical expectation due to the haul length increase, that is $S_\phi(f) \propto L^3$ [5]. Link stability over long periods was calculated by counting the heterodyne beatnote on PD2 with a high resolution Λ-type frequency counter [14]. Data taken with this type of counters leads to a frequency stability that differs from the Allan deviation, in particular in presence of white and flicker phase noise; nevertheless, scaling formulas can be used to convert between the two ones [15]. Figure 4 shows the fractional frequency instability $\sigma_y(\tau)$ of the link at 100 km (circles) and at 200 km (squares); the interferometer floor with no spools inserted (diamonds) is also reported. Fractional frequency instability of a few $10^{-19}$ at 1000 s is obtained at 200 km with one counter-propagating ROA pump, limited by the residual non-reciprocal noise of the optical fiber. ROA is fully reliable for a phase coherent optical link in a counter-propagating pump scheme.

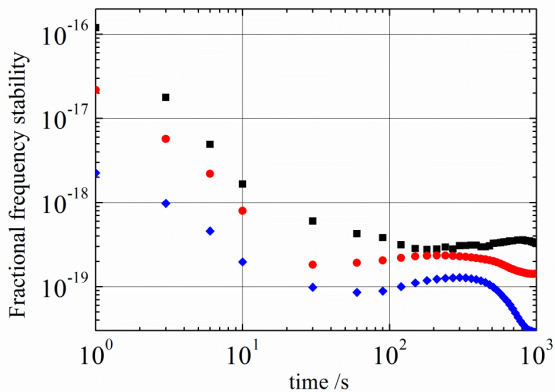

Fig. 4. Fiber link frequency instability using ROA: 100 km (circles), 200 km (squares), interferometer noise floor (diamonds).

To increase the span between intermediate stations, we tested a double stage pump scheme, coupling through DWM1 the Fabry-Perot diode laser as a co-propagating pump. ROA overall gain was increased to 32 dB, equivalent to a single stage ROA with 1 W single counter-propagating pump [11], limited by the diode laser power (260 mW). The double stage pump allowed to implement the link over 61 dB losses, equivalent to a 275 km optical fiber haul, obtained inserting a 16 dB attenuator. This was the maximum length achieved with our setup in a single span. without intermediate amplification.

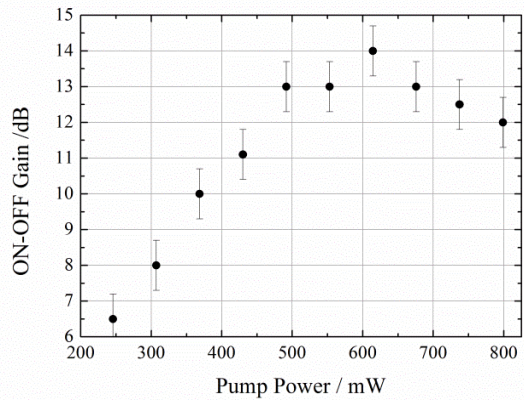

Fig. 5. On-Off Gain in a co-propagating pump scheme, limited by nonlinear effects due to SBS.

We exchanged the two optical pumps, to use the more powerful fiber laser as co-propagating pump. In this case, we observed strong nonlinear effects, due to stimulated Brillouin scattering (SBS) [8], limiting ROA performances. We measured the co-propagating gain keeping the counter-propagating pump at a constant 9 dB gain, while increasing the co-propagating pump power. As shown in Figure 5, because of the nonlinear scattering, the ROA gain saturated in the co-propagating pump, and instead of attaining the previous 23 dB on-off gain, only 13 dB maximum gain was allowed. As the signal power coupled to the optical link was 0.5 mW, we conclude that co-propagating saturation occurs when the signal locally exceeds 10 mW. This should be taken into account considering the design of an optical fiber link requiring intermediate ROA stations.

In conclusion, we demonstrated that ROA is suitable for phase coherent optical links, allowing the optical frequency transfer on 275 km in single fiber span, with a frequency instability of $10^{-19}$ at 1000 s. ROA offers several advantages with respect to other amplification techniques. It does not require pump frequency control, thanks to its wide bandwidth and moreover, using the fiber itself as a gain medium, it has a distributed and higher amplification than the BEDFA systems. However, particular care should be devoted to the design of links involving both counter-propagating and co-propagating pumps, to avoid the triggering of unwanted nonlinear scattering affecting the ultimate noise performances of the link.